\documentclass[twocolumn,showpacs,preprintnumbers,amsmath,amssymb]{revtex4}

\usepackage{graphicx}
\usepackage{dcolumn}
\usepackage{bm}

\begin{document}

\preprint{APS/123-QED}

\title{Liquid Front Profiles Affected by Entanglement-induced Slippage}

\author{O. B\"aumchen}

\author{R. Fetzer}
\altaffiliation{Present address: Ian Wark Research Institute,
University of South Australia, Mawson Lakes, SA 5095, Australia.\\}

\author{K. Jacobs}
 \email{k.jacobs@physik.uni-saarland.de}

 \affiliation{Department of Experimental Physics, Saarland
 University, D-66041 Saarbr\"ucken, Germany.}

\date{\today}

\begin{abstract}
Hydrodynamic slippage plays a crucial role in the flow dynamics of
thin polymer films, as recently shown by the analysis of the
profiles of liquid fronts. For long-chained polymer films it was
reported that a deviation from a symmetric profile is a result of
viscoelastic effects. In this Letter, however, evidence is given
that merely a slip boundary condition at the solid/liquid interface
can lead to an asymmetric profile. Dewetting experiments of
entangled polymer melts on diverse substrates allow a direct
comparison of rim morphologies. Variation of molecular weight $M_w$
clearly reveals that slippage increases dramatically above a certain
$M_w$ and governs the shape of the rim. The results are in
accordance with the theoretical description by de Gennes.
\end{abstract}


\maketitle

The controlled manipulation of small liquid volumes is one of the
main tasks in the field of micro- and nanofluidics. In this context,
the role of hydrodynamic slippage has attracted much attention due
to its enormous relevance for applications such as lab-on-chip
devices. A variety of experimental methods exist to characterize
slippage. The large number of techniques and results is summarized
and discussed in recent review articles \cite{Net05, Lau05, Boc07}.

If a thin liquid film on top of a solid substrate can gain energy by
minimizing its interfacial area, the film ruptures and holes occur
(see \cite{Rei92}, \cite{Jac08} and references therein). These dry
circular patches grow with time, and the removed liquid accumulates
in a rim surrounding the holes along their perimeter. Dewetting of
thin liquid films on solid substrates is a result of internal
capillary forces, which can be inferred from the effective interface
potential \cite{Vri66, See012}. Viscous dissipation within the
liquid, which mainly occurs at the three-phase contact line
\cite{Bro94}, and friction at the solid/liquid interface (slippage)
counteract these driving forces. Concerning the characteristic
dynamics of hole growth (hole radius $R$ versus time $t$),
Brochard-Wyart and colleagues \cite{Bro94} proposed a sequence of
different stages starting with the birth of a hole, which is
characterized by an exponential increase of the dewetting velocity
(denoted by $V=dR/dt$). It is followed by a regime that is governed
by the formation of the rim and dominated by viscous flow ($R\propto
t$), followed by a stage of a ''mature'' rim, where surface tension
rounds the rim and slippage might be involved. In case of full
slippage, a growth law $R\propto t^{2/3}$ is expected. This view was
supplemented by an experimental study by Damman and colleagues, who
associated different dewetting regimes with corresponding rim shapes
in case of viscoelastic fluids \cite{Dam03}. Vilmin $et\,al.$
presented a theoretical approach taking viscoelastic properties and
slippage into account \cite{Vil05}. In our earlier studies, we
experimentally examined rim profiles of dewetting long-chained PS
films on hydrophobized Si wafers and attributed the transition from
symmetric (''oscillatory'') to increasingly asymmetric
(''monotonically decaying'') shapes with the molecular weight of the
polymer to viscoelastic effects \cite{See011}. These results go
along with a phenomenological model that predicts a phase diagram
for rim morphologies \cite{Her02}.

These previous studies, however, lack one facet, which by now has
been identified as one of the major issues influencing dewetting
dynamics and profiles: hydrodynamic slippage. In our recent studies
we focused our interest on slippage of polymer melts, where polymers
were used exclusively below their entanglement length \cite{Fet05,
Fet071, Bae081}. In this Letter, the studies are extended to
long-chain polymer melts. Our aim is to characterize slippage of
polystyrene films above their entanglement length. This might
involve new problems related to non-Newtonian behavior due to
viscoelastic flow and stress relaxation. The latter has been
reported to cause polymer thin film rupture \cite{Pod01, Rei05} and
to impact the early stage of dewetting \cite{Rei05}. As will be
shown in the following for the mature stage, residual stress and
viscoelastic properties have no impact on the shape of dewetting
rims, if low shear rates are involved. It will be demonstrated that
in that stage, asymmetric rim profiles are caused by slippage only.

Films of atactic polystyrene (PSS Mainz, Germany) with molecular
weights ranging from 5.61 to 390~kg/mol (termed e.g. PS(390k)) and
low polydispersity ($M_w/M_n=1.02-1.09$) were spin-cast from toluene
solutions on freshly cleaved mica sheets and transferred to smooth
hydrophobized Si wafers with a native oxide layer (obtained from
Siltronic, Burghausen, Germany) by floating them on a
Millipore\texttrademark water surface. To avoid residual stresses in
films above the entanglement length of PS, the films were
pre-annealed on the mica substrate well above their glass transition
temperature (up to 3 hours at 140$\,^{\circ}$C) \cite{Pod01}. The
prepared film thicknesses $h_0$ varied between 100 and 140\,nm.
Hydrophobization was either achieved through standard silanization
techniques \cite{Was89} using self-assembled monolayers of
octadecyltrichlorosilane (OTS) and dodecyltrichlorosilane (DTS) or
by preparing a thin amorphous teflon layer (AF\,1600) using the
spin-coating technique. The thickness $d$ and the root-mean-square
(\textit{rms}) roughness of the hydrophobic layers (see Tab.
\ref{tab1}) were measured via ellipsometry (EP$^3$, Nanofilm,
G\"ottingen, Germany) and atomic force microscopy (AFM, Multimode,
Veeco Instruments, Santa Barbara, CA, USA). Additionally, substrates
are characterized in terms of their wetting properties by their
contact angles (advancing $\theta_{adv}$, and contact angle
hysteresis $\Delta\theta$) of Millipore\texttrademark water and
their surface energy $\gamma_{sv}$ (via measuring the contact angles
of apolar liquids).

\begin{table}
\caption{\label{tab1}Substrate properties.}
\begin{ruledtabular}
\begin{tabular}{cccccc}
layer & d (nm)& rms (nm)& $\theta_{adv}$ ($^{\circ}$) &
$\Delta\theta$ ($^{\circ}$) & $\gamma_{sv} (mN/m)$ \\
\hline
AF\,1600  &  21(1) &  0.30(3) &  128(2) & 10 & 15.0 \\
OTS  &  2.3(2) &  0.09(1) &  116(1) & 6 & 23.9 \\
DTS  &  1.5(2) &  0.13(2) &  114(1) & 5 & 26.4 \\
\end{tabular}
\end{ruledtabular}
\end{table}

\begin{figure}[t]
\begin{center}
\includegraphics[width=0.4\textwidth]{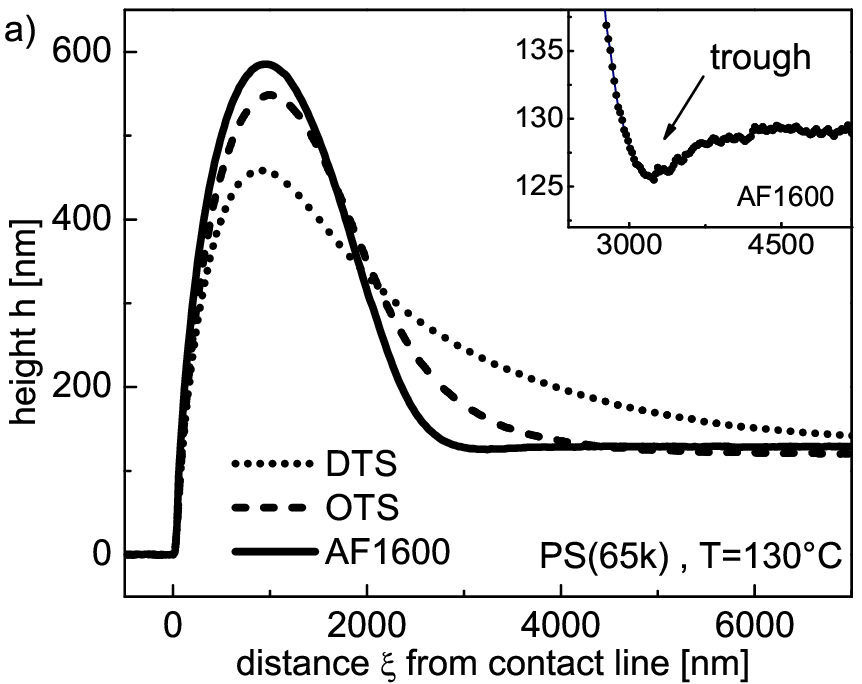}
\includegraphics[width=0.4\textwidth]{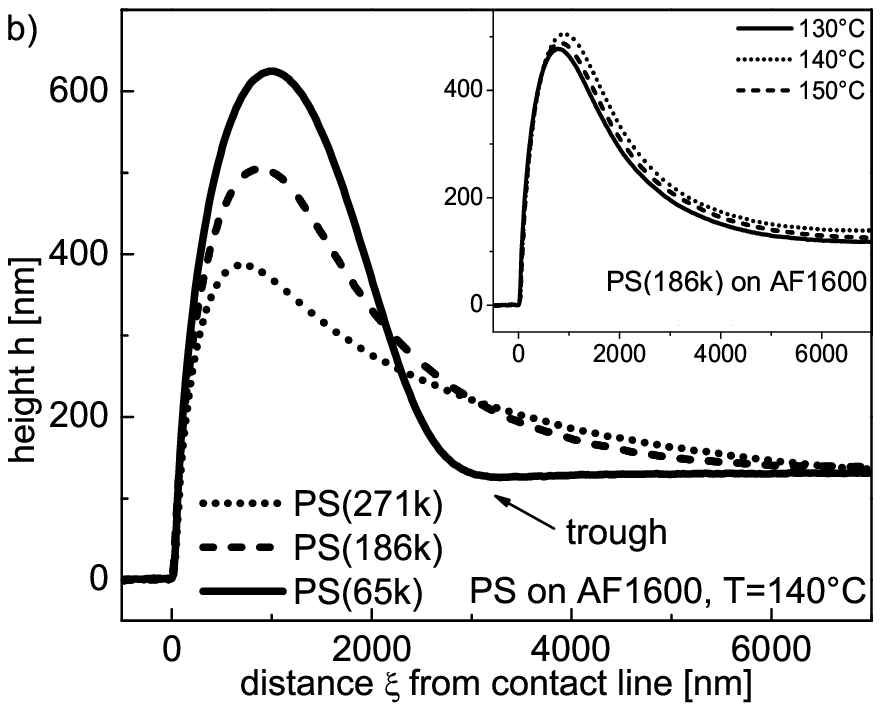}

\end{center}
\caption{a) Rim profiles of PS(65k) on DTS (dotted), OTS (dashed)
and AF\,1600 (solid line) at 130$\,^{\circ}$C. The inset illustrates
the oscillatory behavior on AF\,1600. b) Rim shapes on AF\,1600 for
different molecular weights (140$\,^{\circ}$C, PS(65k), PS(186k),
PS(271k)). Inset: Variation of viscosity via different temperatures
for the same molecular weight (see Tab. \ref{tab2}) shows
insignificant influence on rim profiles.} \label{graphB}
\end{figure}

Nucleation and subsequent growth of holes starts after heating the
samples above the glass transition temperature of the PS melt. In
the following, optical microscopy is utilized to dynamically monitor
the size of growing holes. At a certain hole radius (about
12\,$\mu$m) in the mature regime, the experiment is stopped by
quenching the sample to room temperature. The profile of the rim is
then imaged by AFM \cite{foot1}. As shown in Fig. \ref{graphB} a),
comparing rim profiles for the same molecular weight, film thickness
and viscosity (for the respective dewetting temperature) clearly
exhibits strong differences that can exclusively be ascribed to the
different substrates. For e.g. PS(65k) at 130$\,^{\circ}$C an
oscillatory profile with a ''trough'' on the ''wet'' side of the rim
is observed on the AF\,1600 substrate. A close-up of the trough is
given in the inset. Yet, in the case of OTS and DTS, rims are found
that decay monotonically into the unperturbed polymer film. In Fig.
\ref{graphB} b), experimental profiles are shown for the AF\,1600
substrate, where the molecular weight of the polymer melt is varied.
In case of PS(65k), we observe a clearly oscillatory profile,
whereas for e.g. PS(186k) or higher molecular weights, asymmetric
(monotonically decaying) rim morphologies are recorded. Hence, an
increase in $M_w$ on the identical substrate provokes the same
morphological differences in rim shapes as before (Fig.~\ref{graphB}
a)) the variation of the substrate. To probe whether or not the
effect is merely a result of the melt viscosity, the dewetting
temperature on AF\,1600 (see Tab. \ref{tab2}) was additionally
varied. Yet, no systematic influence of the viscosity on the rim
profiles is found for this substrate, as exemplarily shown for
PS(186k) in the inset of Fig.~\ref{graphB} b). The observed marginal
differences of the rim profiles can be related to small variations
of accumulated material (via slightly different hole radii and film
thickness in different experiments).

Summarizing the qualitative observations, we find that an increase
in molecular weight strongly amplifies the asymmetry of the profiles
in the same manner as a change of substrate is capable of causing.
According to our previous studies on OTS and DTS concerning
non-entangled PS melts, asymmetric rims are related to slippage
\cite{Fet05, Fet071, Bae081}. For DTS, slip lengths $b$ of up to 5
$\mu$m were obtained, for OTS, values are about one order of
magnitude smaller. The question now arises as to the manner in which
slippage correlates to the molecular weight and the number of
entanglements in the AF\,1600 system. To quantify slippage there,
rim profiles will be analyzed.

As shown recently, lubrication models for Newtonian flow and
different slip conditions (no-slip, weak-slip, intermediate-slip and
strong-slip \cite{Mue05}) and a more generalized model based on the
full Stokes equations \cite{Fet071} can be applied to thin dewetting
films with the important result that values for the slip length $b$
can be obtained.

\begin{table}
\caption{\label{tab2}Comparison of viscosities (obtained from the
WLF equation), calculated relaxation times $\tau$ and measured rim
widths $w$ (on AF\,1600) given for a selection (see
Fig.~\ref{graphB}) of dewetting temperatures $T$ and molecular
weights $M_w$.}
\begin{ruledtabular}
\begin{tabular}{ccccc}
$M_w$ & $T$ ($^{\circ}$C)& $\eta$ (Pa s)& $\tau$ (s) & $w$ ($\mu$m)\\
\hline
 65k & 130  &  $1.3 \times 10^6$ & 6.5  & 2.73  \\
 65k & 140  & $1.6 \times 10^5$ & 0.8  & 2.81  \\
 186k & 130  & $4.4 \times 10^7$ & 222  & 5.02  \\
 186k & 140  & $5.6 \times 10^6$ & 28.1  & 4.86  \\
 186k & 150  & $1.1 \times 10^6$ & 5.4  & 5.29  \\
 271k & 140  & $2.0 \times 10^7$ & 100.5  & 7.47  \\
\end{tabular}
\end{ruledtabular}
\end{table}

To assess whether these models might be used to analyze the
experimental data, the impact of viscoelasticity on fluid flow needs
to be estimated: The Weissenberg number $Wi=\tau\dot{\gamma}$
relates the time scale of stress relaxation $\tau$ with the time
scale of applied shear $\dot\gamma$. Relaxation times are determined
from the corresponding viscosities $\eta$ and the shear modulus
$G\approx0.2\,$MPa (for PS, \cite{Rub03}) via $\tau=\eta/G$ (see
Tab. \ref{tab2}). Evaluation of the shear rates within the moving
rim is based on $\dot\gamma = max(\partial_x u_x,\partial_z
u_x)\approx max(\dot s/\Delta x,\dot s/\Delta z)$, where $\dot{s}$
is the dewetting speed (identified as the dominant velocity $u_x\gg
u_z$ within the rim) which can be inferred from optical hole radius
measurements. The lateral displacement $\Delta x$ is estimated as
the width $w$ of the rim. The value $w$ is defined as the distance
between the three-phase contact line and the position where the rim
height has dropped to 110\% of the film thickness $h_0$, i.e.
$h(w)=1.1h_0$. The vertical displacement $\Delta z$ is given by
$\Delta z=h_0+b$, where $b$ denotes the slip length as determined by
the quantitative rim analysis described in the last part of this
Letter. The described estimation leads to values for $Wi$ between
0.01 and 0.1 (on AF\,1600) and 0.5 (on OTS/DTS). An influence of
viscoelastic effects on flow dynamics is expected for Weissenberg
numbers of 1 and larger. In that case, a model including
viscoelasticity has to be used \cite{Blo06}. Low shear rates as
described above and the pre-annealing step after spin-coating allow
us to safely exclude viscoelastic effects as the source of different
rim morphologies. This consideration justifies the assumption of
Newtonian flow, on which the following data analysis is based.

The theoretical description of the model for Newtonian liquids
presented in \cite{Fet071} is derived from the full Stokes equations
in two dimensions for a viscous, incompressible liquid including the
Navier slip-boundary condition $b=u/\partial_z u|_{z=0}$, where $u$
denotes the velocity in $x$-direction and $z$ the vertical scale.
This leads to the equations of motion for a flat liquid film. Then,
a linear stability analysis is applied by introducing a small
perturbation $\delta h= h(x,t)-h_0$ and velocity $u(x,t)$ to the
undisturbed state $h=h_0$ and $u=0$. In the co-moving frame
$\xi=x-s(t)$, where $s(t)$ is the position of the rim, a
quasi-stationary profile evolves. Solving the linear equation using
the normal modes ansatz $\delta h=\delta h_0 \exp(k\xi)$ and $u=u_0
\exp(k\xi)$ gives a characteristic equation for $k$ which is
expanded up to third order according to the Taylor formalism:

\begin{equation}
\label{chareq}
(1+\frac{h_0}{3b})(h_0k)^3+4Ca(1+\frac{h_0}{2b})(h_0k)^2-Ca\frac{h_0}{b}=0,
\end{equation}

\noindent where $Ca$ denotes the capillary number given by $Ca=\eta
\dot{s}/\gamma_{sl}$. Hence, rim profiles depend on the capillary
number $Ca$ and the ratio of slip length $b$ to film thickness
$h_0$. Eq. (\ref{chareq}) implies that a morphological transition
from oscillatory (complex conjugate solutions for $k$) to
monotonically (real $k$ values) decaying rim shapes occurs if
$Ca^2>(3^3(b/h_0+1/3)^2)/(4^4(b/h_0+1/2)^3)$ on slippery substrates
(large $b/h_0$) and for large capillary numbers $Ca$ \cite{Fet071,
Bae081}. Furthermore, Eq. (\ref{chareq}) predicts progressively
asymmetric rims for increasing values of $b/h_0$.

\begin{figure}[t]
\begin{center}
\includegraphics[width=0.45\textwidth]{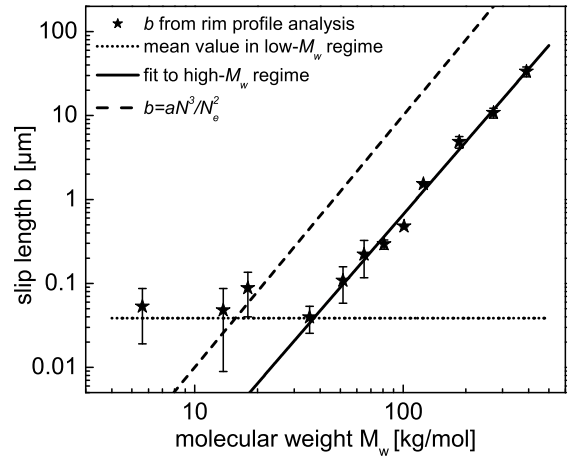}
\end{center}
\caption{Slip lengths of PS films on AF\,1600 obtained from rim
profile analysis. Each data point represents the average value of
different measurements (2-3 measurements for each $M_w$ value above
35.6k and 5 or more below). The solid line represents a fit to the
experimental data from PS(35.6k) to PS(390k), the dashed line
visualizes $b= aN^3/N^2_e$ using values for $a$ and $N_e$ given in
the literature \cite{Rub03}. The dotted line represents the average
of slip lengths for melts from PS(5.61k) to PS(18k).} \label{graphC}
\end{figure}

In the following, PS melts on the AF\,1600 substrate are considered:
Rim profiles are evaluated by fitting the wet side of the rim with a
damped oscillation (oscillatory profiles, up to PS(81k)) or a simple
exponential decay (monotonic rims, for PS(101k) and above).

In the case of an exponential decay, a single value for $k$ is
gained from the experimental data. Knowing the film thickness $h_0$
and the capillary number $Ca$ (from independent rim velocity data),
this $k$ value allows us to calculate the respective slip length via
Eq. (\ref{chareq}). For oscillatory profiles, $k$ is a pair of
complex conjugate numbers. From these two independent solutions,
both the capillary number and the slip length can be extracted via
Eq. (\ref{chareq}) at the same time. Viscosity data calculated from
these $Ca$ values showed excellent agreement with theoretical values
(from the WLF equation) and demonstrate the consistency of the
applied model. For further details concerning the fitting procedure
and data evaluation, one should refer to previous publications
\cite{Fet05, Fet071, Bae081}.

The slip lengths $b$ of a set of experiments, i.e. different
dewetting temperatures and viscosities, are shown in
Fig.~\ref{graphC} as a function of molecular weight $M_w$ of the PS
melt. Two distinct regimes can clearly be seen: Constant slip
lengths in the order of 10 to 100\,nm at small $M_w$ and slip
lengths that increase with the power of 2.9(2) in case of larger
$M_w$. The transition between both regimes occurs at 37\,kg/mol,
which is consistent with the critical value $M_c$ of entanglement
effects ($M_c$=35\,kg/mol for PS according to \cite{Rub03}). For the
same viscosity (e.g. 2~$\times$~10$^6$~Pa\,s) in case of e.g.
PS(51.5k) and PS(271k), an enormous difference in slippage of two
orders of magnitude solely due to the difference in chain length is
obtained. We therefore state that slippage is directly related to
entanglements rather than viscosity. Besides the fact that an
influence of the melt viscosity (via temperature variation) for a
certain molecular weight could not be detected, even experiments
omitting the pre-annealing step after preparation on mica do not
show significant impact. Additionally, we want to emphasize that
careful analysis of the dewetting velocity data from hole growth
dynamics (of the identical samples), according to the model
presented in \cite{Fet072}, confirmed the $b(M_w)$ behavior obtained
from rim analysis \cite{Bae09}.

For long polymer chains, described by the reptation model, de Gennes
predicts a scaling of $b\propto N^3$, where $N$ denotes the number
of monomers, which corroborates the experimental exponent of 2.9(2)
for experiments with $M_w>$~35~kg/mol as depicted in
Fig.~\ref{graphC}. According to the model, $b= aN^3/N^2_e$, where
$N_e$ is the bulk entanglement length (number of monomers in an
entanglement strand) and $a$ a polymer specific molecular size, and
is plotted in Fig.~\ref{graphC} as the dashed line for the
literature values of $N_e=163$ \cite{Rub03} and $a=3{\AA}$
\cite{Red94} for PS. From the linear fit to our experimental data
(solid line in Fig.~\ref{graphC}), we obtain
$a/N_e^2$=1.12(7)$\times$10$^{-5}${\AA}, which is one order of
magnitude smaller that expected by using the literature values given
above. Keeping $a$ from the literature, a significantly larger
entanglement length of $N_e=517$ results from the linear fit. This
is an indication for a significantly lowered \textit{effective}
entanglement density in the vicinity of the substrate compared to
the bulk and is in line with recent studies of thin polymer films
\cite{Si05,Bar09}. According to Brown and Russell, $N_e$ of a
polymer melt near an interface is expected to be about 4 times that
in the bulk \cite{Bro96}, which corroborates our experimental
results.

To conclude, entanglements can strongly amplify slippage as deduced
from dewetting experiments. Assigning asymmetric profiles
exclusively to non-Newtonian effects such as viscoelasticity
therefore turns out to be invalid. On the contrary, the hydrodynamic
boundary condition, i.e. slippage, can dramatically provoke
morphological changes in rim shape such as transitions from
oscillatory to monotonic profiles. The onset of slippage correlates
to the critical chain length for entanglements. The dependence of
the slip length on molecular weight corroborates the description by
de Gennes. To derive vice versa slip lengths solely from de Gennes'
theory entails that bulk values for $a$ and $N_e$ (from literature)
have to be used that might not appropriately describe polymer melts
next to an interface. The method described here, however, represents
a powerful tool to gain experimentally access to liquid flow
properties at the solid/liquid interface and to quantify slip
lengths.

We gratefully acknowledge financial support from German Science
Foundation DFG under Grant Ja 905/3 within the priority programm
1164 and generous support of Si wafers by Wacker Siltronic AG,
Burghausen, Germany.


\bibliography{prl}

\begin{thebibliography}{99.}
\bibitem{foot1} Note that quenching the sample to room temperature has been proven not to change or
influence the shape of the rim profile.
\end{thebibliography}

\end{document}